\documentclass[times,preprint,a4paper,11pt]{elsarticle}
\usepackage[verbose=true,a4paper]{geometry}
\usepackage[bookmarks=false]{hyperref}
    \hypersetup{colorlinks,
      linkcolor=blue,
      citecolor=blue,
      urlcolor=blue}

\makeatletter
\def\ps@pprintTitle{%
 \let\@oddhead\@empty
 \let\@evenhead\@empty
 \def\@oddfoot{\centerline{\thepage}}%
 \let\@evenfoot\@oddfoot}
\makeatother

\usepackage{amssymb}
\usepackage{amstext}
\usepackage{amsmath}
\usepackage{amsthm}
\usepackage{xcolor}
\usepackage{tikz}

\usepackage{enumerate} 
\usepackage{algorithm}
\usepackage[noend]{algpseudocode}
\usepackage{hyperref} 
\usepackage{url,caption,subcaption}
\usepackage{color,bm}
\usepackage{soul} 
\usepackage{makecell} 
\usepackage{array}
\newcolumntype{C}[1]{>{\centering\let\newline\\\arraybackslash\hspace{0pt}}m{#1}}
\newcolumntype{L}[1]{>{\raggedright\let\newline\\\arraybackslash\hspace{0pt}}m{#1}}
\newcolumntype{R}[1]{>{\raggedleft\let\newline\\\arraybackslash\hspace{0pt}}m{#1}}
\usepackage{booktabs} 
\usepackage{nicematrix}
\makeatletter
\def\@citex[#1]#2{%
  \let\@citea\@empty
  \@cite{\@for\@citeb:=#2\do
    {\@citea\def\@citea{;\penalty\@m\ }%
     \edef\@citeb{\expandafter\@firstofone\@citeb}%
     \if@filesw\immediate\write\@auxout{\string\citation{\@citeb}}\fi
     \@ifundefined{b@\@citeb}{\mbox{\reset@font\bfseries ?}%
       \G@refundefinedtrue
       \@latex@warning
         {Citation `\@citeb' on page \thepage \space undefined}}%
       {\csname b@\@citeb\endcsname}}}{#1}}
\makeatother
\usepackage{tabularx, environ}
\makeatletter
\newcolumntype{\expand}{}
\long\@namedef{NC@rewrite@\string\expand}{\expandafter\NC@find}
\NewEnviron{problem}[2][]{%
  \def\problem@arg{#1}%
  \def\problem@framed{framed}%
  \def\problem@lined{lined}%
  \def\problem@doublelined{doublelined}%
  \ifx\problem@arg\@empty%
    \def\problem@hline{}%
  \else%
    \ifx\problem@arg\problem@doublelined%
      \def\problem@hline{\hline\hline}%
    \else%
      \def\problem@hline{\hline}%
    \fi%
  \fi%
  \ifx\problem@arg\problem@framed%
    \def\problem@tablelayout{|>{\bfseries}lX|c}%
    \def\problem@title{\multicolumn{2}{|l|}{%
        \raisebox{-\fboxsep}{\textsc{\normalsize #2}}%
      }}%
  \else
    \def\problem@tablelayout{>{\bfseries}lXc}%
    \def\problem@title{\multicolumn{2}{l}{%
        \raisebox{-\fboxsep}{\textsc{\normalsize #2}}%
      }}%
  \fi%
  \bigskip\par\noindent%
  \begin{tabularx}{\textwidth}{\expand\problem@tablelayout}%
    \problem@hline%
    \problem@title\\[2\fboxsep]%
    \BODY\\\problem@hline%
  \end{tabularx}%
  \medskip\par%
}
\makeatother

\usepackage{pgfplots}
\usepackage{graphicx}

\def\ra{{\rightarrow}}

\def\bF{{\mathbb F}}
\def\bN{{\mathbb N}}
\def\bZ{{\mathbb Z}}

\def\cI{{\mathcal I}}

\def\cM{{\mathcal M}}

\def\cO{{\mathcal O}}

\def\cV{{\mathcal V}}

\newcommand{\Hw}[1]{\left|#1\right|}
\newcommand{\supp}[1]{\mathrm{Supp}(#1)}

\newcommand\norm[1]{\left\lVert#1\right\rVert} 

\def\mA{\bm{A}}

\def\m0{\bm{0}}

\def\vx{\bm{x}}

\def\v0{\bm{0}}

\newtheorem{remark}{Remark}

\newtheorem{proposition}{Proposition}

\newtheorem{theorem}{Theorem}

\begin{document}

\begin{frontmatter}

\title{On the Low Weight Polynomial Multiple Problem}

\author[a]{Ferucio Lauren\c{t}iu \c{T}iplea\corref{cor1}} 
\ead{ferucio.tiplea@uaic.ro}
\author[b]{Simona-Maria L\u{a}z\u{a}rescu}
\ead{lazarescusimonam@gmail.com}

\address[a]{Faculty of Computer Science, 
						``Alexandru Ioan Cuza'' University of Ia\c si, 
						Ia\c si, Romania}
\address[b]{``Simion Stoilow'' Institute of Mathematics of the Romanian Academy,
            Bucharest, Romania}

\begin{abstract}
Finding a low-weight multiple (LWPM) of a given polynomial is very useful 
in the cryptanalysis of stream ciphers and arithmetic in finite fields. 
There is no known deterministic polynomial time complexity algorithm for 
solving this problem, and the most efficient algorithms are based on a 
time/memory trade-off. The widespread perception is that this problem is 
difficult. In this paper, we establish a relationship between the 
LWPM problem and the MAX-SAT problem of determining an assignment that 
maximizes the number of valid clauses of a system of affine Boolean clauses. 
This relationship shows that any algorithm that can compute the optimum 
of a MAX-SAT instance can also compute the optimum of an equivalent 
LWPM instance. It also confirms the perception that the LWPM problem 
is difficult.
\end{abstract}

\begin{keyword}
Optimization problem \sep low-weight polynomial multiple \sep maximum satisfiability

\end{keyword}
\cortext[cor1]{Corresponding author.}
\end{frontmatter}

\section{Introduction}

The Low-Weight Polynomial Multiple (LWPM) problem consists of finding 
a polynomial of degree at most $n$ and minimum Hamming weight to be 
a multiple of a given polynomial $P$ of some degree $d$. 
There are various variants of this problem, 
such as determining a multiple polynomial of $P$ of degree at most $n$ 
and weight at most $w$ or determining the set of all multiple 
polynomials of $P$ of degree at most $n$ and weight at most $w$. 
This problem, which has applications in the cryptanalysis of stream 
ciphers \cite{Sieg1985,MeSt1989,JoJo2002,ChJM2002,Meie2011,TIMAZ2018,MaJJ2023} 
and arithmetic in finite fields \cite{GaNo2005}, 
proved to be quite difficult. All attempts to solve it efficiently 
(through deterministic algorithms of polynomial time complexity) have failed.
Even if the general perception is that this problem is difficult, a proof 
in this direction is missing.

The techniques proposed so far to approach the LWPM problem fall into four classes: 
time/memory trade-off (TMTO) techniques \cite{Goli1996,ChJM2002,Aima2021}, 
techniques based on discrete logarithm \cite{DiLa2007,PeST2016}, 
strategies that reduce the problem to code theory problems \cite{LoJo2014},
and lattice-based techniques \cite{AiGa2007}.

In \cite{Goli1996}, the author proposes a technique that works in two 
phases. The first phase generates a list of polynomials of the form 
$x^{i_1}+\cdots+x^{i_j}\bmod P(x)$, where $0\leq i_i<\cdots<i_j<n$. 
Then, the second phase uses a sort-and-match procedure to determine 
a collision. Any collision gives rise to a multiple polynomial of 
$P$ of weight $2j$. The time and space complexity is about 
$\cO\left(\binom{n}{j}\right)$.

The approach in \cite{ChJM2002} is match-and-sort, looking for collisions 
in a space of size $n^w$. The time complexity of the algorithm is 
$\cO(n^{w/2}\log{n})$, and the space complexity is $\cO(n^{(w-1)/4})$.

Viewing the solution to an LWPM instance as a collision, \cite{Aima2021}
proposes two memory-efficient algorithms. Then, using the parallel
collision search technique in \cite{OoWi1999}, the author tune the two
algorithms to decrease the running time at the expense of memory. 

\cite{DiLa2007} proposes a solution based on the calculation of 
discrete logarithms whose complexity is $\cO(n^{w/2})$. 
Later, following the same approach based on discrete logarithms, 
\cite{PeST2016} provided a memory-efficient algorithm that runs 
in $\cO(2^d/n)$. However, the method assumes a constant cost of 
calculating the discrete logarithm, which is not valid if $2^d-1$ 
is not smooth. Both \cite{DiLa2007} and \cite{PeST2016} require 
other restrictive conditions on the polynomial $P$.

An attractive solution was proposed in \cite{AiGa2007}. The authors 
view polynomials as vectors, structure the set of multiple polynomials 
of $P$ as a lattice, construct a basis for this lattice, and then 
use the LLL algorithm to reduce the basis. In this way, solutions 
for the LWPM problem could be obtained. The time complexity of this 
solution will be $\cO(n^6)$ (dictated by the complexity of the LLL 
algorithm), and the space $\cO(nd)$. This method gives 
inaccurate results as soon as the bound $n$ exceeds few hundreds. 

\cite{LoJo2014} proposes an algorithm for finding low-weight multiples 
of polynomials over the binary field using coding theoretic methods. 
The associated code defined by the given polynomial has a cyclic 
structure, allowing an algorithm to search for shifts of the sought 
low-weight codeword. Thus, the LWPM problem is reduced to the low-weight 
codeword problem, which is also difficult.

\paragraph{Contributions}
In this paper, we connect the LWPM problem and the MAX-SAT problem, 
which asks for a Boolean assignment that maximizes the number of 
valid affine constraints in a given system of such constraints. For 
this, we view the polynomials as Toeplitz matrices. We thus show that 
the LWPM problem can be strictly reduced to the MAX-SAT problem. This 
means that any optimal solution for a MAX-SAT instance will produce 
an optimal solution for the original LWPM instance. As a result, 
through this reduction, any algorithm for solving the MAX-SAT problem 
can also be applied to the LWPM problem. Conversely, we can only 
highlight a probabilistic reduction from the MAX-SAT problem to the 
LWPM problem. The experimental analysis shows us that, for sufficiently 
large instances, this reduction produces MAX-SAT solutions whose measure 
is approximately equal to the measure of the LWPM solutions from which 
they are built.

\paragraph{Paper organization}
The paper is structured into five sections. The next one recalls basic 
concepts and notations that are used throughout the paper. In Section 3
we view the LWPM problem as an optimization one and we prove several 
basic properties on it. In Section 4 we related the LWPM and MAX-SAT 
problems. This, we show first the the LWPM problem can strctly be
reduced to the MAX-SAT problem, and then we consider the converse
reduction. We conclude the paper in the last section.

\section{Preliminaries}\label{sec:Prelim}

In this section, we will briefly recall the basic concepts used in the paper.
For details, the reader is referred to 
\cite{Lang2002,DuFo2003,Gent2017,ACGKMP2003, CrKS2001}. 

\paragraph{Polynomials and matrices}

Generic fields are denoted by $\bF$. When we want to emphasize 
that a field is finite and has the order $q$, we will write 
$\bF_q$. $\bF^n$ stands for the $n$-dimensional vector space 
over $\bF$. The vectors of $\bF^n$ will be denoted by lowercase 
letters, such as $\vx$, and written in column form. 
The $i$th element of $\vx\in\bF^n$ is denoted $\vx(i)$, 
where $1\leq i\leq n$, and the support of $\vx$ is  
$\supp{\vx}=\{i\mid 1\leq i\leq n,\,\vx(i)\not =0\}$.
The cardinality of the support of $x$ is the Hamming weight of $\vx$.
We shall simply denote this as $\Hw{\vx}$. 
The operator ``$|\cdot|$'' is used both for the Hamming weight and the 
cardinality of a set. However, the distinction will always be clear 
from the context.  

A polynomial in the indeterminate $x$ over a field $\bF$ is written 
as a formal sum $P(x)=a_0+a_1x+\cdots+a_nx^n$, where $n\geq 0$ and
each $a_i$ is from $\bF$. If $a_n\not=0$, then $P$ is said to be of 
degree $n$, denoted $deg(P)=n$. 
The weight of $P$ is the Hamming weight of the vector $(a_0,a_1,\ldots,a_n)$. 

The set of $m\times n$ matrices with elements in $\bF$ is denoted 
$\cM_{m,n}(\bF)$.For a matrix $A$, $A(i,j)$ denotes the element of $\mA$ at 
the intersection of row $i$ and column $j$ and $A^t$ is the transpose 
of $A$.

\paragraph{Optimization problems}

An \textit{optimization problem} \cite{ACGKMP2003, CrKS2001} is a tuple $A=(\cI,Sol,\mu,Opt)$,
where:
\begin{itemize}
\item $\cI$ is a set of \textit{instances};
\item $Sol$ is a function that maps each instance $x$ to a set
	of feasible solutions of $x$;
\item $\mu$ is a \textit{measure function} that associates to each pair
 	$(x,y)\in\cI\times Sol(x)$ a positive integer regarded as a measure 
 	of the instance $x$'s feasible solution $y$;
\item $Opt\in\{min,max\}$ is the \textit{optimization criterion} (minimization
  	or maximization). 
\end{itemize}

Given an optimization problem as above, we denote by $Sol^*(x)$ the
set of optimal solutions of $x$ (that is, those feasible solutions 
from $Sol(x)$ that fulfill the optimization criterion $Opt$). 
Moreover, $\mu^*(x)$ stands for the value of the optimal solutions
of $x$, when they exist. 

Each optimization problem $A$ defines three related problems:
\begin{itemize}
\item \textit{Constructive problem} $A_C$: Given an instance $x\in\cI$, 
  compute an optimal solution and its measure;
\item \textit{Evaluation problem} $A_E$: Given an instance $x\in\cI$,
  compute $\mu^*(x)$;
\item \textit{Decision problem} $A_D$: Given an instance $x\in\cI$ and 
  a positive integer $t$, decide the predicate
  \begin{center}
		$
    A_D(x,t)=\left\{
               \begin{array}{ll}
                 1, & (Opt=min\ \wedge\ \mu^*(x)\leq t)\ \vee\ \\ [.5ex]
                    & (Opt=max\ \wedge\ \mu^*(x)\geq t) \\ [.5ex] 
                 0, & \text{otherwise.}
               \end{array}
             \right.
		$
  \end{center}
\end{itemize}
The \textit{underlying language} of $A$ is the set
$\{(x,t)\in\cI\times\bN\mid A_D(x,t)=1\}$. 

An optimization problem $A=(\cI,Sol,\mu,Opt)$ belongs to the class
NPO if:
\begin{itemize}
\item Its set of instances is recognizable in polynomial time;
\item There is a polynomial $p$ such that for each instance $x$,
  	its feasible solutions have size at most $p(|x|)$. Moreover,
  	for each $y$ of size at most $p(|x|)$, it is decidable in polynomial
  	time whether $y$ is a feasible solution of $x$;
\item The measure function is computable in polynomial time. 
\end{itemize}

Let $A=(\cI_A,Sol_A,\mu_A,Opt_A)$ and $B=(\cI_B,Sol_B,\mu_B,Opt_B)$ 
be two NPO problems. A \textit{reduction} from $A$ to $B$ is a pair 
$(f,g)$ of polynomial-time computable function such that:
\begin{itemize}
\item $f(x)\in\cI_B$, for any $x\in\cI_A$;
\item $g(x,y)\in Sol_A(x)$, for any $x\in\cI_A$ and $y\in Sol_B(f(x))$. 
\end{itemize}

The reduction $(f,g)$ is called \textit{strict} or an \textit{$S$-reduction} 
from $A$ to $B$ if the following two properties hold:
\begin{itemize}
\item $\mu_B^*(f(x))=\mu_A^*(x)$, for any $x\in\cI_A$;
\item $\mu_A(x,g(x,y))=\mu_B(f(x),y)$, for any $x\in\cI_A$ and 
  	$y\in Sol_B(f(x))$. 
\end{itemize}

\section{The low-weight polynomial multiple problem}

Finding a polynomial of low weight but which is a multiple of a given 
polynomial $P$ proved to be very important in the cryptanalysis of 
stream ciphers and finite field arithmetic. 
Formally, the problem is as follows. Let $P(x)\in\bF[x]$ be a polynomial
over a commutative ring $\bF$ and $n$ and $w$ be strictly positive integers. 
The problem asks to find a polynomial $K(x)\in\bF[x]$ of degree less
than $n$ and weight at most $w$ such that $P\mid K$. 
Some authors strengthen the requirement by finding a polynomial of 
the lowest weight that is a multiple of $P$, and others require finding 
all polynomials of the lowest weight that are multiples of $P$. 

Working in a commutative ring provides generality, but the size of 
the feasible solutions may be arbitrarily large. For example, the 
following polynomials are multiples of the polynomial $P(x)=1+x+x^2\in\bZ[x]$
\begin{center}
  $
  \begin{array}{lcl}
    P(x)(2-2x) &=& 2-2x^3 \\
    P(x)(2-2x)(2+2x^3) &=& 4-4x^6 \\
    P(x)(2-2x)(2+2x^3)(2+2x^6) &=& 8-8x^{12}
  \end{array}
  $
\end{center}
It is clear how new polynomials with arbitrarily large coefficients 
can be defined.
We also notice that even if we work in a finite field, the degree of 
the solutions (the multiples of $P$) can be arbitrarily large.

Due to practical aspects of computational complexity, we will impose 
two requirements on the problem:
\begin{enumerate}
\item The polynomial $P$ in any instance $(P,n,w)$ of the problem is over a 
  finite field $\bF_q$;
\item The parameter $n$ in any instances $(P,n,w)$ of the problem is 
  polynomial in the degree of the polynomial $P$, that is, 
  $n=\cO(deg(P)^k)$ for some $k\geq 1$.
\end{enumerate}

Analyzing this problem carefully, we find that it takes a decisional 
form (by which it is asked to decide if there is such a multiple), 
a constructive form (by which it is required to find such a multiple),
and also a form of evaluation of the minimum weight for which there is 
such a multiple. 
As a result, it is natural to formulate the problem as an optimization 
problem.

\begin{problem}[framed]{MIN Polynomial Multiple (MIN-PM) Problem}
Instance: & Polynomial $P\in\bF[x]$ and positive integers $n>deg(P)$
            of polynomial size in $deg(P)$, where $\bF$ is a finite field; \\
Sol:      & Non-zero polynomial $K\in\bF[x]$ of degree less that $n$ such that 
            $P\mid K$; \\
Measure:  & Hamming weight of polynomials (number of non-zero coefficients); \\
Opt:      & $min$.  
\end{problem}

The decision problem associated with the MIN-PM problem asks, given a 
polynomial $P$ and two integers $n$ and $w$, to decide if there is a 
multiple $K$ of $P$, of degree less than $n$ and weight at most $w$. 
The constructive problem associated with the MIN-PM problem asks, given 
a polynomial $P$ and an integer $n$, to find a multiple $K$ of $P$ of 
degree at most $n$ and lowest weight. Both the decision problem and 
the constructive one associated with MIN-PM are among the problems 
encountered so far in the literature. To our knowledge, the evaluation 
problem associated with the MIN-PM problem has not been 
mentioned so far in the literature. This is probably because what 
interests us the most is the computation of an optimal measure solution, 
and finding such a solution immediately leads to the optimal value 
of the measure.

We define the size of a MIN-PM instance $(P,n)$ as the size of its 
binary representation, i.e.,
\begin{center} 
  $\norm{(P,n)}=(d+1)\log{q}+\log{n},$ 
\end{center}
where $deg(P)=d $ and $q$ is the number of elements of the finite field $\bF$. 
As $n=\cO(d^k)$ for some $k\geq 1$, we can rewrite the instance size as
\begin{center}
  $\norm{(P,n)}=(d+1)\log{q}+k\log{d}.$
\end{center}

\begin{proposition}
The following properties hold:
\begin{enumerate}
\item MIN-PM is an $NPO$ problem. 
\item $MIN\text{-}PM_D\equiv_T^p MIN\text{-}PM_E\leq_T^p MIN\text{-}PM_C$.
\item $MIN\text{-}PM_D\in NP$. 
\end{enumerate}
\end{proposition}
\begin{proof}
(1) We will check the membership requirements of the MIN-PM problem 
for the NPO class.
First, we remark that any instance of MIN-PM can be recognized in 
polynomial time with respect to its size.  

Let $(P,n)$ be a MIN-PM instance, $d=deg(P(x))$, $k\geq 1$ such that
$n=\cO(d^k)$, and $p(x)$ be the polynomial $p(x)=x^k$.
 
If $K(x)$ is a feasible solution to $(P,n)$, then 
\begin{center}
  $\norm{K(x)}=n\log{q}=\cO(d^k\log{q})=\cO(p(\norm{(P,n)})).$
\end{center}
If $K(x)\in\bF[x]$ is a polynomial of size $p(\norm{(P,n)})$, then one
can check in polynomial time whether $K(x)$ is a feasible solution to
$(P,n)$.

Finally, the measure function is computible in linear time with respect to
the instance size (just inspect the polynomial coefficients). 

(2) and (3) follow directly from (1) (see Theorems 1.1 and 1.4 in 
\cite{ACGKMP2003}).
\end{proof}

\section{MIN-PM and MAX-SAT problems}

We will establish in this section a connection between the MIN-PM and 
MAX-SAT problems. To work as easily as possible from a technical point 
of view, we will restrict the study to the case $\bF=\bF_2$. 
The extension to finite fields can easily be obtained. 

Let us first recall the MAX-SAT problem \cite{KoKM1994,CrKS2001}.
Consider $D$ a finite set of cardinality at least two that we 
fix for the rest of the paper.
A \textit{constraint} over a finite set $\cV$ of variables is a pair 
$\sigma=(f,(x_{1},\ldots,x_{\ell}))$ consisting of a function 
$f:D^\ell\ra\{0,1\}$, for some $\ell\geq 1$, and a list 
$(x_{1},\ldots,x_{\ell})$ of variables from $\cV$ that 
take values in $D$. 
The function $f$ is called the \textit{constraint function}, and the
list of variables, the \textit{constraint scope} of $\sigma$. 
The variables in the constraint scope may not be pairwise distinct. 

An \textit{instance of the constraint satisfaction problem} over a set
$\cV$ of variables is a finite set $\Sigma$ of constraints over $\cV$.
Distinct constraints in $\Sigma$ may have distinct constraint functions 
or scopes.

Let $\gamma:\cV\ra D$ be an assignment of the variables in $\cV$.
A constraint $\sigma=(f,(x_{1},\ldots,x_{\ell}))$ over $\cV$ is
\textit{satisfied} by $\gamma$ is $f(\gamma(x_1),\ldots,\gamma(x_\ell))=1$.

Now, we are able to formulate the following problem. 

\begin{problem}[framed]{MAX Constraint Satisfaction Problem (MAX-CSP)}
Instance: & Finite set $\Sigma$ of constraints over a finite set $\cV$
            of variables; \\
Goal:     & Assignment $\gamma:\cV\ra D$; \\
Measure:  & Number of constraints in $\Sigma$ satisfied by  $\gamma$; \\
Opt:      & $max$.  
\end{problem}

When $D=\{0,1\}$ and each constraint function is a Boolean function,
MAX-CSP is usually referred to as the MAX-SAT problem. When the
constraints are affine, that is, expressions of the form
\begin{center}
   $x_1\oplus\cdots\oplus x_n=b,$
\end{center}
where $n\geq 1$, $x_1,\cdots,x_n\in\cV$, and $b\in\{0,1\}$, the problem 
is usually denoted MAX-SAT(affine).

\subsection{From MIN-PM to MIN-SAT}

Determining a multiple $K(x)$ of minimum weight of a polynomial $P(x)$ 
is equivalent to determining a polynomial $Q(x)$ so that the polynomial 
$P(x)Q(x)$ is of minimum weight. If we represent the polynomials by 
their vectors of coefficients, the coefficients of the product $P(x)Q(x)$ 
of the polynomials $P(x)$ and $Q(x)$ can be obtained by multiplying 
a Toeplitz matrix $M_{P,t}$ associated with $P (x)$ with the column vector 
associated with $Q(x)$.
The Toeplitz matrix $M_{P,t}$ has $d+t+1$ rows and $t+1$ columns, where 
$P(x)=a_0+a_1x+\cdots+a_dx^d$ and $Q(x)=b_0+b_1x+\cdots+b_tx^t$. 
The first column is $(a_0,a_1,\ldots,a_d,0,\ldots,0)^t.$
The other columns are obtained by cyclically permuting the first column 
down by one position. For example, the coefficients of the polynomial 
$P(x)Q(x)$, where $P(x)=1+x+x^2$ and $Q(x)=1-x+x^3-x^4$ are obtained by 
multiplying $M_{P,4}$ (on the left) with the column vector associated to
$Q$ (on the right). 
\begin{center}
$
\begin{pmatrix}
1 & 0 & 0 & 0 & 0 \\
1 & 1 & 0 & 0 & 0 \\
1 & 1 & 1 & 0 & 0 \\
0 & 1 & 1 & 1 & 0 \\
0 & 0 & 1 & 1 & 1 \\
0 & 0 & 0 & 1 & 1 \\
0 & 0 & 0 & 0 & 1
\end{pmatrix}
\cdot
\begin{pmatrix}
1 \\
-1 \\ 
0 \\
1 \\
-1
\end{pmatrix}
$
\end{center}

\begin{figure}
\centering
\begin{tikzpicture}[font=\scriptsize]
\draw[fill=blue!3]  (-2.5,1.8) node (v3) {} ellipse (1.1 and 1);
\draw[fill=blue!3]  (2.5,1.8) node (v3) {} ellipse (1.1 and 1);
\node (x) at (-2.5,1.75) {$(P,n)$};
\node (x') at (2.5,1.75) {$M_{P,t}\cdot x_Q=1$};
\draw[-latex] (x) .. controls (-0.5,2.25) and (0.5,2.25) .. node[above] {$f$} (x');
\draw[fill=gray!10] (2.5,1.5) node (v1) {} -- (1.5,-1) -- (3.5,-1) -- (v1);
\draw[fill=gray!10] (-2.5,1.5) node (v2) {} -- (-3.5,-1) -- (-1.5,-1) -- (v2);
\node (y') at (2.5,0) {$\gamma$};
\node (y) at (-2.5,0) {$P\cdot Q$};
\draw[-latex] (y') .. controls (0.5,-0.5) and (-0.5,-0.5) .. node[above] {$g$} (y);
\node at (-2.5,-1.25) {$Sol((P,n))$};
\node at (2.5,-1.25) {$Sol(M_{P,t}\cdot x_Q=1)$};
\node at (-2.5,3) {MIN-PM};
\node at (2.5,3) {MIN-SAT};
\end{tikzpicture}
\caption{The reduction $(f,g)$ in Theorem \ref{T-Red_to_MAX}}
\label{F1}
\end{figure}
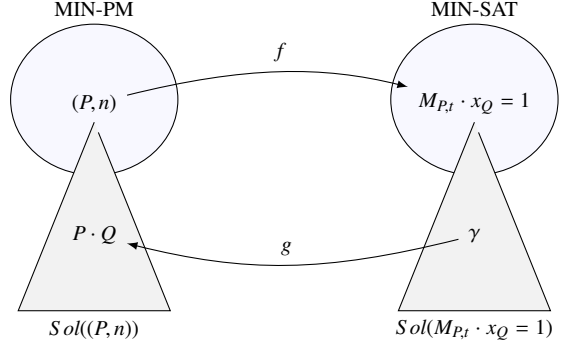

\begin{theorem}\label{T-Red_to_MAX}
MIN-PM$(\bF_2)$ is $S$-reducible to MAX-SAT(affine).
\end{theorem}
\begin{proof}
Consider the following reduction $(f,g)$ from MIN-PM$(\bF_2)$ to MIN-SAT(affine):
\begin{enumerate}
\item Given a MIN-PM$(\bF_2)$ instance $(P,n)$, $f((P,n))$ is the 
  MIN-SAT(affine) instance ``$M_{P,t}\cdot x_Q=0$'', where $n=deg(P)+t+1$ and
  $x_Q=(x_0,\ldots,x_t)^t$ is a vector of Boolean variables;
\item Given a solution $\gamma$ to ``$M_{P,t}\cdot x_Q=0$'', $g((P,n),\gamma)$
  is the polynomial $K(x)=P(x)\cdot Q(x)$, where 
  $Q(x)=\gamma(x_0)+\gamma(x_1)x+\cdots+\gamma(x_t)x^t$. 
\end{enumerate}

The reduction is pictorially represented in Figure \ref{F1}.
Due to the fact that $n$ is polynomial in the degree of $P$, the instance
``$M_{P,t}\cdot x_Q=0$'' can be computed in polynomial time in the size
of $(P,n)$. Similarly, $P(x)\cdot Q(x)$ can be computed in polynomial time
in the size of $\gamma$. Therefore, the reduction $(f,g)$ is polynomial-time
computable. 

If $\mu_{PM}$ and $\mu_{SAT}$ are the measure functions of MIN-PM and
MAX-SAT, respectively, then one can easily check that the following properties 
hold:
\begin{itemize}
\item $\mu_{PM}((P,n),g((P,n),\gamma))=\mu_{SAT}(f((P,n)),\gamma)$; 
\item $\mu_{SAT}^*(f((P,n)))=\mu_{PM}^*((P,n))$.
\end{itemize}
Therefore, $(f,g)$ is an $S$-reduction. 
\end{proof}

\begin{remark}
The reduction of MIN-PM to MAX-SAT is optimal. What we have to do 
is just to construct the Toeplitz matrix of the polynomial given in the
LWPM instance. Any optimal solution returned by the MAX-SAT solver is 
optimal for the LWPM instance. 
Therefore, we obtain in this way a SAT solver for MIN-PM 
(see Algorithm \ref{MIN-PM-Alg1}). 
\end{remark}

\begin{algorithm}[!h]
    \caption{SAT solver for MIN-PM$(\bF_2)$}
    \label{MIN-PM-Alg1} 
    \begin{algorithmic}[1]
        \Function{MIN-PM\_solve}{$(P,n)$} 
        \State Compute the Toeplitz matrix $M_{P,t}$, where 
               $n=deg(P)+t+1$;
        \State Compute the MAX-SAT(affine) instance 
           ``$M_{P,t}\cdot x_Q=0$'', where $x_Q=(x_0,\ldots,x_t)^t$;
        \State $\gamma:=$MAX-SAT\_{\scriptsize SOLVE}$(M_{P,t}\cdot x_Q=0)$;
        \State Compute $Q(x)=\gamma(x_0)+\gamma(x_1)x+\cdots+\gamma(x_t)x^t$;
        \State\textbf{return} $P(x)Q(x)$. 
        \EndFunction
    \end{algorithmic}
\end{algorithm}

\subsection{From MAX-SAT to MIN-PM}

The reduction from the previous section shows us that MAX-SAT is 
heavier than MIN-PM. We believe they would be just as heavy, but 
unfortunately, we didn't manage to make an S-reduction from MAX-SAT 
to MIN-PM. We still managed to establish a \textit{probabilistic reduction} 
from MAX-SAT to MIN-PM. It works like this:
\begin{enumerate}
\item Given a matrix $A$ that defines a MAX-SAT instance, we construct 
	a Teoplitz form of it, $A_T$;
\item $A_T$ defines a polynomial $P$ and thus a MIN-PM instance;
\item Starting from a solution $P(x)Q(x)$ for the instance $A_T$, through 
	a probabilistic algorithm we build a solution $\gamma$ for the 
	instance $A$;
\item The experimental analysis shows us that the measure of the 
	solution $\gamma$ is smaller than the measure of the solution $P(x)Q(x)$, 
	except for a negligibly small number of cases.
\end{enumerate}

\medskip
For the experiments, we used two genetic algorithms, more precisely 
\textit{Stochastic Hill Climbing} and \textit{Simulated Annealing}. 
Genetic algorithms \cite{Mitc1998,Schm2001,Cham2000,HaHa2004,Carr2014} 
are optimization algorithms inspired by natural selection and genetics. 

The Hill Climbing algorithm (see Algorithm \ref{Hill-Climbing_Alg}) 
is a meta-heuristic that, starting from an initial solution, tries to 
improve it gradually, iteratively, until no better solution is found. 
There are several types of such algorithms, but among the most used 
are \textit{Simple Hill Climbing}, in which the state of the neighboring 
node at a particular moment is evaluated, \textit{Steepest Ascent Hill Climbing}, 
in which all neighbors are taken into account and is selected the 
best node, and \textit{Stochastic Hill Climbing}, where the neighbors 
are chosen randomly. 

\begin{algorithm}[!h]
    \caption{Hill Climbing}
    \label{Hill-Climbing_Alg} 
    \begin{algorithmic}[1]
        \Function{Hill\_climbing}{initial\_solution} 
        \State current\_solution=initial\_solution
        \Repeat
            \State neighbours = generate\_neighbours(current\_solution)
            \State best\_neighbours = best(neighbours)
            \State random\_neighbor = random(best\_neighbours)
            \If{fitness(random\_neighbor) $\leq$ fitness(current\_solution)}
                \State current\_solution=random\_neighbor
                \State \textbf{break}. 
            \EndIf
        \Until {fitness(current\_solution) $\leq$ fitness(of all its neighbours)}
        \EndFunction
    \end{algorithmic}
\end{algorithm}

Simulated Annealing (see Algorithm \ref{Simulated-Annealing_Alg})
is also a meta-heuristic, which originates in the 
physical annealing process from real life, where a material is heated 
up until an annealing temperature is reached, and then a gradual 
cooling process is carried out with the aim of changing the state of 
the material to a certain desired state. 

\begin{algorithm}[!h]
    \caption{Simulated Annealing}
    \label{Simulated-Annealing_Alg} 
    \begin{algorithmic}[1]
        \Function{Simulated\_Annealing}{initial\_solution,T} 
        \State current\_solution=initial\_solution
        \While {true}
             \If {T $\leq T_{min}$}
                \State \textbf{return} current\_solution.
            \EndIf
            \State Select neighbor $\in$ generate\_neighbours(current\_solution)
            \If{fitness(neighbor) $\leq$ fitness(current\_solution)}
                    \State current\_solution=random\_neighbor
            \ElsIf{random() $<$ exp((fitness(current\_solution) - fitness(neighbor))/T)}
                \State current\_solution=random\_neighbor
            \EndIf
            \State T=T$\times$alpha
        \EndWhile
        \EndFunction
    \end{algorithmic}
\end{algorithm}

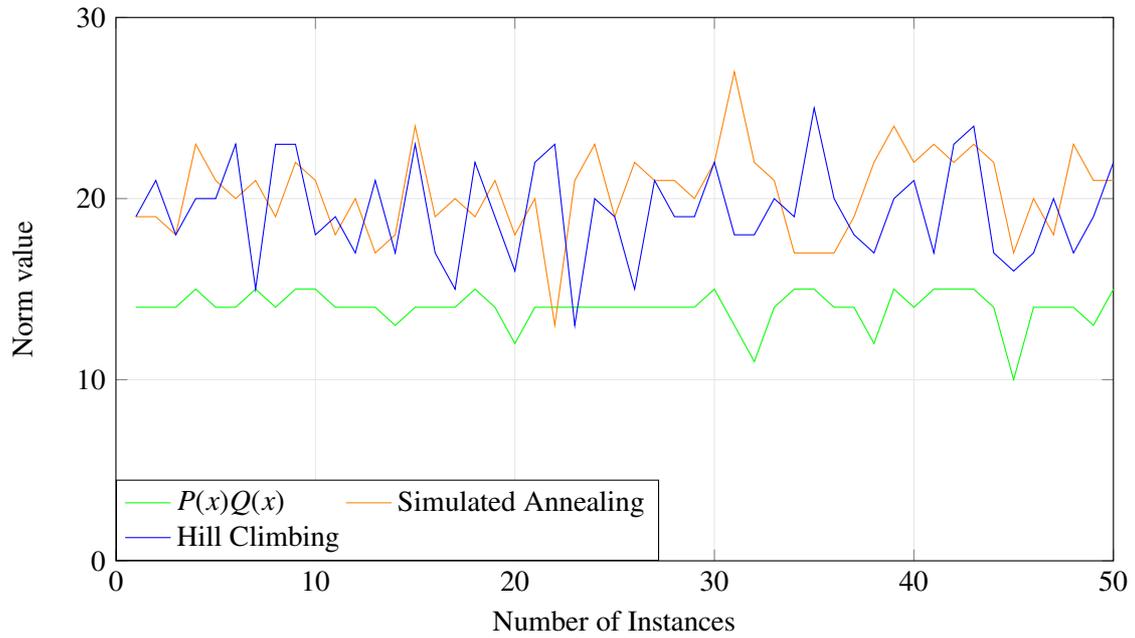
\begin{figure}[hbt]
    \centering
             \begin{tikzpicture}
            \begin{axis}[
            width=\textwidth,
            height=250pt,
    legend columns=2,
    legend style={at={(0.0,0.0)},anchor=south west,legend cell align=left},
    ylabel=Norm value,
    xlabel=Number of Instances,
    ymin=0,
    xmin=0,
    xmax=50,ymax=30,
    xtick={0,10, 20, 30, 40, 50},
    ytick={0,10, 20, 30, 35},
    grid=both,
    grid style={line width=.1pt, draw=gray!20}]
                \addplot+[mark=none,green] table [x, y, col sep=comma] {grafice/40_30_pq.txt};\addlegendentry{$P(x)Q(x)$}
                \addplot+[mark=none,orange] table [x, y, col sep=comma] {grafice/40_30_sa.txt};\addlegendentry{Simulated Annealing}
                \addplot+[mark=none,blue] table [x, y, col sep=comma] {grafice/40_30_hc.txt};\addlegendentry{Hill Climbing}
                
                \end{axis}
         \end{tikzpicture}
    \caption{Probabilistic reduction from MAX-SAT to MIN-PM for 
             matrices of size $40\times 30$}
    \label{Fig-40x30}
\end{figure}

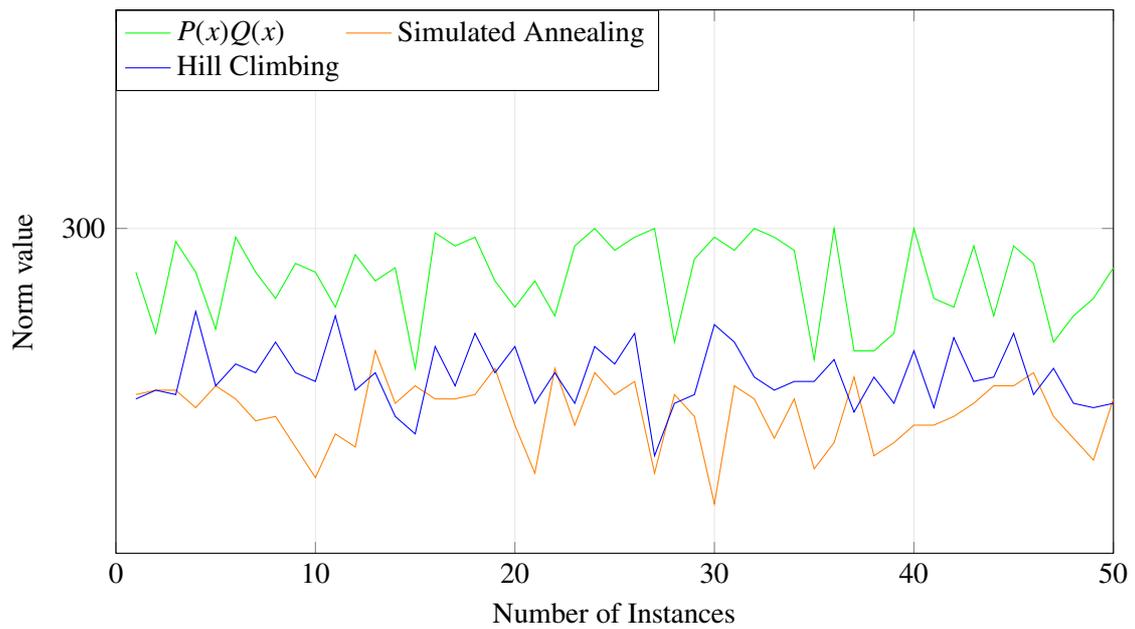
\begin{figure}[!ht]
    \centering
             \begin{tikzpicture}
            \begin{axis}[
            width=\textwidth,
            height=250pt,
    legend columns=2,
    legend style={at={(0.0,1.0)},anchor=north west,legend cell align=left},
    ylabel=Norm value,
    xlabel=Number of Instances,
    xmin=0,
    xmax=50,ymax=350,
    xtick={0,10, 20, 30, 40, 50},
    ytick={0,100, 200, 300, 400},
    grid=both,
    grid style={line width=.1pt, draw=gray!20}]
                \addplot+[mark=none,green] table [x, y, col sep=comma] {grafice/400_200_pq.txt};\addlegendentry{$P(x)Q(x)$}
                \addplot+[mark=none,orange] table [x, y, col sep=comma] {grafice/400_200_sa.txt};\addlegendentry{Simulated Annealing}
                \addplot+[mark=none,blue] table [x, y, col sep=comma] {grafice/400_200_hc.txt};\addlegendentry{Hill Climbing}
                
                \end{axis}
         \end{tikzpicture}
    \caption{Probabilistic reduction from MAX-SAT to MIN-PM for 
             matrices of size $400\times 200$}
    \label{Fig-400x200}
\end{figure}

\begin{figure}[!ht]
    \centering
             \begin{tikzpicture}
            \begin{axis}[
            width=\textwidth,
            height=250pt,
    legend columns=2,
    legend style={at={(1.0,1.0)},anchor=north east,legend cell align=left},
    ylabel=Norm value,
    xlabel=Number of Instances,
    xmin=0,
    xmax=50,ymax=750,
    xtick={0,10, 20, 30, 40, 50},
    ytick={0,100, 200, 300, 400, 500, 600, 700, 800},
    grid=both,
    grid style={line width=.1pt, draw=gray!20}]
                \addplot+[mark=none,green] table [x, y, col sep=comma] {grafice/1000_500_pq.txt};\addlegendentry{$P(x)Q(x)$}
                \addplot+[mark=none,orange] table [x, y, col sep=comma] {grafice/1000_500_sa.txt};\addlegendentry{Simulated Annealing}
                \addplot+[mark=none,blue] table [x, y, col sep=comma] {grafice/1000_500_hc.txt};\addlegendentry{Hill Climbing}
                
                \end{axis}
         \end{tikzpicture}
    \caption{Probabilistic reduction from MAX-SAT to MIN-PM for 
             matrices of size $1000\times 500$}
    \label{Fig-1000x500}
\end{figure}
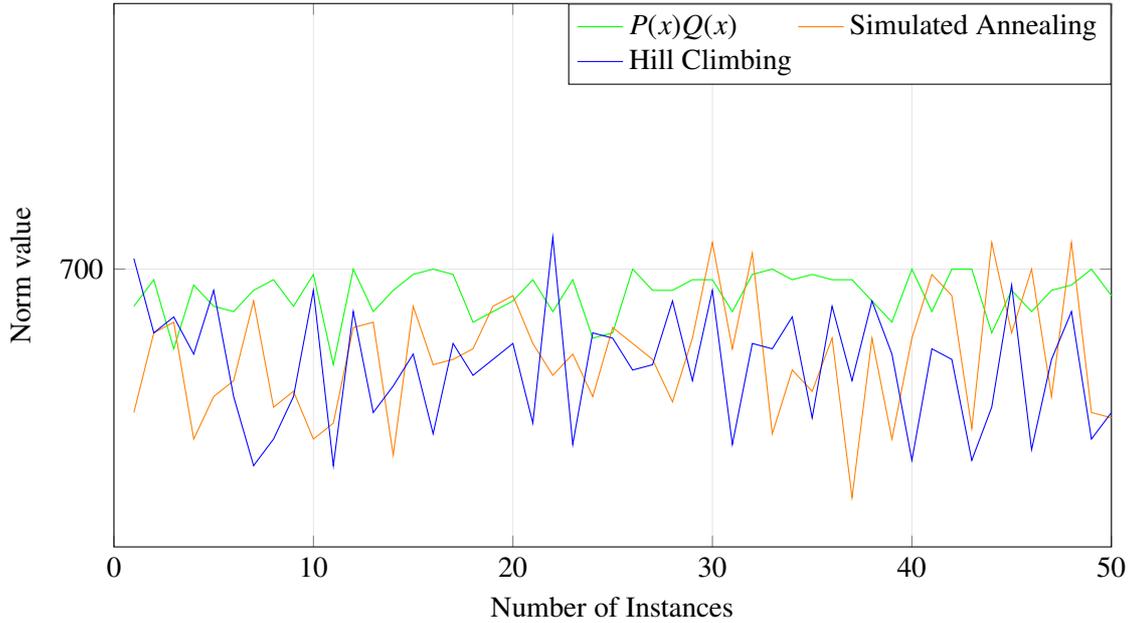

We have considered instances $A$ for MAX-SAT of type $40\times 30$, 
$400\times 200$, and $1000\times 500$. 
For each instance, the Toeplitz matrix $A_T$ defining a polynomial 
$P$ was built. A multiple $P(x)Q(x)$ of $P(x)$ was derived. Then, using 
the Hill Climbing and Simulated Annealing algorithms a solution $\gamma$
was generated for the instance $A$. We compared the norm of the 
solution $\gamma$ with the norm of the solution $P(x)Q(x)$. 
The results are graphed below (Figures \ref{Fig-40x30}, \ref{Fig-400x200},
\ref{Fig-1000x500}): 
the norm of $P(x)Q(x)$ is represented in green, 
the norm of $\gamma$ obtained by the Hill Climbing algorithm is represented
in blue, and the norm of $\gamma$ obtained by the Simulated Annealing algorithm 
is represented in orange.
 
Experimentally, we found that the ratio between the norm of the solution 
$\gamma$ and the norm of $P(x)Q(x)$ is very close to 1 for instances 
greater than $400\times 200$.
Considering the S-reduction from Section 4.1, which shows that a MIN-PM 
instance is reduced to a MAX-SAT instance for which the optimum is preserved, 
we believe that for large instances the reduction suggested by the 
probabilistic algorithms in this section would produce instances $\gamma$ 
with optimum very close to the optimum of MIN-PM instances. Thus, we believe 
that this reduction shows experimentally that MIN-PM is as hard as MAX-SAT.
 
The results are systematized in the Table \ref{table:Results}.

\begin{table}
\centering
\begin{tabular}{|c|c|c|l|}
\hline
MAX-SAT instance & LWPM instance & Max ratio for HC & Max ratio for SA \\ \hline
40$\times$30 & 71$\times$31 & 1.6666666666666667 & 2.076923076923077 \\ \hline
400$\times$200 & 601$\times$201 & 0.9929078014184397 & 0.9850746268656716 \\ \hline
1000$\times$500 & 1501$\times$501 & 1.0202312138728324 & \multicolumn{1}{c|}{1.0247093023255813} \\ \hline
\end{tabular}
\caption{Systematization results.}
\label{table:Results}
\end{table}

\section{Conclusion}
In this paper, we established the connection between the problem of 
determining a low-weight multiple (LWPM) of a given polynomial and the
 MAX-SAT problem. We viewed the LWPM problem as an optimization problem 
called MIN-PM. We established some properties of the MIN-PM problem 
and showed that it can efficiently be S-reduced to the MAX-SAT problem. 
The S-reduction guarantees the optimality of the solution. As a result, 
any algorithm for MAX-SAT can be efficiently converted into an 
algorithm for MIN-PM. Conversely, we showed that MAX-SAT probabilistically 
reduces to MIN-PM. Even if, from a theoretical point of view, the 
probabilistic reduction still needs to be investigated in depth, 
the experimental results suggest that the MIN-PM problem is at least 
as hard as the MAX-SAT problem.

\section*{Authors contribution}
The study of the connection between the MIN-PM and MAX-SAT problems 
using Toeplitz matrices was proposed by F.L. \c{T}iplea. He carried out
Sections 3 and 4.1 (the reduction from MIN-PM to MAX-SAT) and proposed the 
reduction of the MAX-SAT problem to MIN-PM through Toeplitz matrices. 
The use of probabilistic algorithms for achieving the reduction (Section 4.2) 
was proposed by S.-M. L\u{a}z\u{a}reascu, who also performed the experimental 
analysis (see items 3 and 4 of the methodology described at the 
beginning of Section 4.2). 
The rest of the paper was prepared with equal 
contributions from F. L. \c{T}iplea and S.-M. L\u{a}z\u{a}reascu. 



\begin{thebibliography}{}

\bibitem[Aimani and von~zur Gathen, 2007]{AiGa2007}
Aimani, L.~E. and von~zur Gathen, J. (2007).
\newblock Finding low weight polynomial multiples using lattices.
\newblock {\em {IACR} Cryptol. ePrint Arch.}, page 423.

\bibitem[Ausiello et~al., 2003]{ACGKMP2003}
Ausiello, G., Crescenzi, P., Gambosi, G., Kann, V., Marchetti-Spaccamela, A.,
  and Protasi, M. (2003).
\newblock {\em Complexity and Approximation: Combinatorial Optimization
  Problems and Their Approximability Properties}.
\newblock Springer-Verlag, Berlin, Heidelberg, 2nd corrected printing edition.

\bibitem[Carr, 2014]{Carr2014}
Carr, J. (2014).
\newblock An introduction to genetic algorithms.
\newblock {\em Senior Project}, 1(40):7.

\bibitem[Chambers, 2000]{Cham2000}
Chambers, L.~D. (2000).
\newblock {\em The practical handbook of genetic algorithms: applications}.
\newblock Chapman and Hall/CRC.

\bibitem[Chose et~al., 2002]{ChJM2002}
Chose, P., Joux, A., and Mitton, M. (2002).
\newblock Fast correlation attacks: An algorithmic point of view.
\newblock In Knudsen, L.~R., editor, {\em Advances in Cryptology --- EUROCRYPT
  2002}, pages 209--221, Berlin, Heidelberg. Springer Berlin Heidelberg.

\bibitem[Creignou et~al., 2001]{CrKS2001}
Creignou, N., Khanna, S., and Sudan, M. (2001).
\newblock {\em Complexity Classifications of Boolean Constraint Satisfaction
  Problems}.
\newblock Society for Industrial and Applied Mathematics, USA.

\bibitem[Didier and Laigle-Chapuy, 2007]{DiLa2007}
Didier, F. and Laigle-Chapuy, Y. (2007).
\newblock Finding low-weight polynomial multiples using discrete logarithm.
\newblock In {\em 2007 IEEE International Symposium on Information Theory},
  pages 1036--1040.

\bibitem[Dummit and Foote, 2003]{DuFo2003}
Dummit, D. and Foote, R. (2003).
\newblock {\em Abstract Algebra}.
\newblock Graduate Texts in Mathematics. Wiley, 3rd edition.

\bibitem[El~Aimani, 2021]{Aima2021}
El~Aimani, L. (2021).
\newblock A new approach for finding low-weight polynomial multiples.
\newblock In Yu, Y. and Yung, M., editors, {\em Information Security and
  Cryptology}, pages 151--170, Cham. Springer International Publishing.

\bibitem[Gentle, 2017]{Gent2017}
Gentle, J.~E. (2017).
\newblock {\em Matrix Algebra. Theory, Computations and Applications in
  Statistics}.
\newblock Springer Texts in Statistics. Springer International Publishing.

\bibitem[Goli\'{c}, 1996]{Goli1996}
Goli\'{c}, J.~D. (1996).
\newblock Computetion of low-weight parity-check polynomials.
\newblock {\em Electronics Letters}, 32:1979--1980.

\bibitem[Haupt and Haupt, 2004]{HaHa2004}
Haupt, R.~L. and Haupt, S.~E. (2004).
\newblock {\em Practical genetic algorithms}.
\newblock John Wiley \& Sons.

\bibitem[J"{o}nsson and Johansson, 2002]{JoJo2002}
J"{o}nsson, F. and Johansson, T. (2002).
\newblock A fast correlation attack on lili-128.
\newblock {\em Information Processing Letters}, 81(3):127--132.

\bibitem[Kohli et~al., 1994]{KoKM1994}
Kohli, R., Krishnamurti, R., and Mirchandani, P. (1994).
\newblock The minimum satisfiability problem.
\newblock {\em SIAM J. Discrete Math.}, 7:275--283.

\bibitem[Lang, 2002]{Lang2002}
Lang, S. (2002).
\newblock {\em Algebra}.
\newblock Graduate Texts in Mathematics. Springer New York, 3rd edition.

\bibitem[L{\"o}ndahl and Johansson, 2014]{LoJo2014}
L{\"o}ndahl, C. and Johansson, T. (2014).
\newblock Improved algorithms for finding low-weight polynomial multiples in
  $f_2[x]$ and some cryptographic applications.
\newblock {\em Designs, Codes and Cryptography}, 73:625 -- 640.

\bibitem[Ma et~al., 2023]{MaJJ2023}
Ma, S., Jin, C., and Guan, J. (2023).
\newblock Fast correlation attacks on k2 stream cipher.
\newblock {\em IEEE Transactions on Information Theory}, 69(8):5426--5439.

\bibitem[Meier, 2011]{Meie2011}
Meier, W. (2011).
\newblock Fast correlation attacks: Methods and countermeasures.
\newblock In Joux, A., editor, {\em Fast Software Encryption}, pages 55--67,
  Berlin, Heidelberg. Springer Berlin Heidelberg.

\bibitem[Meier and Staffelbach, 1989]{MeSt1989}
Meier, W. and Staffelbach, O. (1989).
\newblock Fast correlation attacks on certain stream ciphers.
\newblock {\em J. Cryptol.}, 1(3):159?176.

\bibitem[Mitchell, 1998]{Mitc1998}
Mitchell, M. (1998).
\newblock {\em An introduction to genetic algorithms}.
\newblock MIT press.

\bibitem[Oorschot and Wiener, 1999]{OoWi1999}
Oorschot, P. C.~v. and Wiener, M.~J. (1999).
\newblock Parallel collision search with cryptanalytic applications.
\newblock {\em Journal of Cryptology}, 12:1--28.

\bibitem[Peterlongo et~al., 2016]{PeST2016}
Peterlongo, P., Sala, M., and Tinnirello, C. (2016).
\newblock A discrete logarithm-based approach to compute low-weight multiples
  of binary polynomials.
\newblock {\em Finite Fields and Their Applications}, 38:57--71.

\bibitem[Schmitt, 2001]{Schm2001}
Schmitt, L.~M. (2001).
\newblock Theory of genetic algorithms.
\newblock {\em Theoretical Computer Science}, 259(1-2):1--61.

\bibitem[Siegenthaler, 1985]{Sieg1985}
Siegenthaler, T. (1985).
\newblock Decrypting a class of stream ciphers using ciphertext only.
\newblock {\em IEEE Transactions on Computers}, C-34(1):81--85.

\bibitem[Todo et~al., 2018]{TIMAZ2018}
Todo, Y., Isobe, T., Meier, W., Aoki, K., and Zhang, B. (2018).
\newblock Fast correlation attack revisited: Cryptanalysis on full {Grain-128a,
  Grain-128, and Grain-v1}.
\newblock In {\em Advances in Cryptology -- CRYPTO 2018: 38th Annual
  International Cryptology Conference, Santa Barbara, CA, USA, August 19-23,
  2018, Proceedings, Part II}, page 129?159, Berlin, Heidelberg.
  Springer-Verlag.

\bibitem[von~zur Gathen and N{\"o}cker, 2005]{GaNo2005}
von~zur Gathen, J. and N{\"o}cker, M. (2005).
\newblock Polynomial and normal bases for finite fields.
\newblock {\em Journal of Cryptology}, 18:337--355.

\end{thebibliography}

\end{document}